\documentclass[superscriptaddress,pra,aps,twocolumn,10pt,longbibliography]{revtex4-1}

\usepackage{amsfonts}
\usepackage{amssymb}
\usepackage{amsmath}
\usepackage{graphicx}
\usepackage{verbatim}
\usepackage{hyperref}
\usepackage[english]{babel}
\usepackage{natbib}
\usepackage[pdftex]{color}
\usepackage[dvipsnames]{xcolor}

\hypersetup{
    colorlinks=true,       
    linkcolor=cyan,          
    citecolor=magenta,        
    filecolor=magenta,      
    urlcolor=cyan,           
    runcolor=cyan
}

\graphicspath{{figs/}}

\newcommand{\fn}[1]{\setcitestyle{open={[[},close={]]}}\textbf{\footnote{#1}}\setcitestyle{square}\hspace{-0.125cm}}

\newcommand{\citedfn}[1]{\setcitestyle{open={[[},close={]]}}\textbf{\cite{#1}}\setcitestyle{square}\hspace{-0.125cm}}

\begin{document}

\title{Distillation of Indistinguishable Photons}

\author{Jeffrey Marshall}\thanks{jmarshall@usra.edu}
\affiliation{QuAIL, NASA Ames Research Center, Moffett Field, CA 94035, USA}
\affiliation{USRA Research Institute for Advanced Computer Science, Mountain View, CA 94043, USA}

\begin{abstract}
A reliable source of  identical (indistinguishable) photons is a prerequisite for exploiting interference effects, which is a necessary component for linear optical based quantum computing, and applications thereof such as Boson sampling.
Generally speaking, the degree of distinguishability will determine the efficacy of the particular approach, for example by limiting the fidelity of constructed resource states, or reducing the complexity of an optical circuits output distribution.
It is therefore of great practical relevance to engineer heralded sources of highly {pure and} indistinguishable photons.
Inspired by magic state distillation, we present a protocol using standard linear optics which can be used to increase the indistinguishability of a photon source, to arbitrary accuracy. 
In particular, in the asymptotic limit of small error $\epsilon$, to reduce the error to $\epsilon' < \epsilon$ requires $O((\epsilon/\epsilon')^2)$ photons. 
{We demonstrate the scheme is robust to detection and control errors in the optical components, and discuss the effect of other error sources.}
\end{abstract}

\maketitle

\textit{Introduction--}
Linear optical quantum computing (LOQC) is an attractive paradigm for realizing fault-tolerance, since photons in free space have extremely long coherence times, and can be manipulated via high fidelity linear optics which may not require the same level of cooling as other approaches \cite{LOQC-review-2007}. In LOQC, qubits are constructed out of photons which can exist in two modes, common choices being spatial modes, or using the polarization degrees of freedom.
Fault tolerance can in principle be achieved via the {Knill-Laflamme-Milburn (KLM)} protocol with sufficient numbers of qubits and using error correction \cite{KLM}, or using cluster states in a measurement-based approach to quantum computing \cite{one-way-QC, deterministic-LOQC, nielsen-opticalQC, one-way, mbqc,LOQC-review-2007,fusion-based}.

{A source of highly indistinguishable photons is required in order to make use of
photons for computational purposes.}
The Hong-Ou-Mandel (HOM) effect \cite{HOM} is the prototypical example which shows fundamental differences in which identical versus distinguishable photons interfere (or do not).
In this conceptually simple experiment, two photons are incident upon a 50:50 beamsplitter, which results in a bunching of the two photons in the case they are indistinguishable.
On the other hand, when the input photons are distinguishable, the signal from an HOM experiment (the HOM `dip') is diminished by an amount related to the infidelity of the two photons \cite{hom-distinguishable}. 

The HOM effect is a crucial ingredient for realizing LOQC, for the interference between identical photons can be used to create entanglement over computational degrees of freedom \cite{KLM, ghz, bleeding, RALPH2010209}. For example, fusion measurements can be used to create large cluster states out of primitive entangled states, such as Bell states or small {Greenberger–Horne–Zeilinger (GHZ)} states \cite{browne-rudolph}. However, the presence of distinguishability will generally result in less entanglement generated over the computational degrees of freedom, compared to the ideal state \cite{sparrow-thesis, birchall-thesis}.

\begin{figure}
    \centering
    \includegraphics[width=0.98\columnwidth]{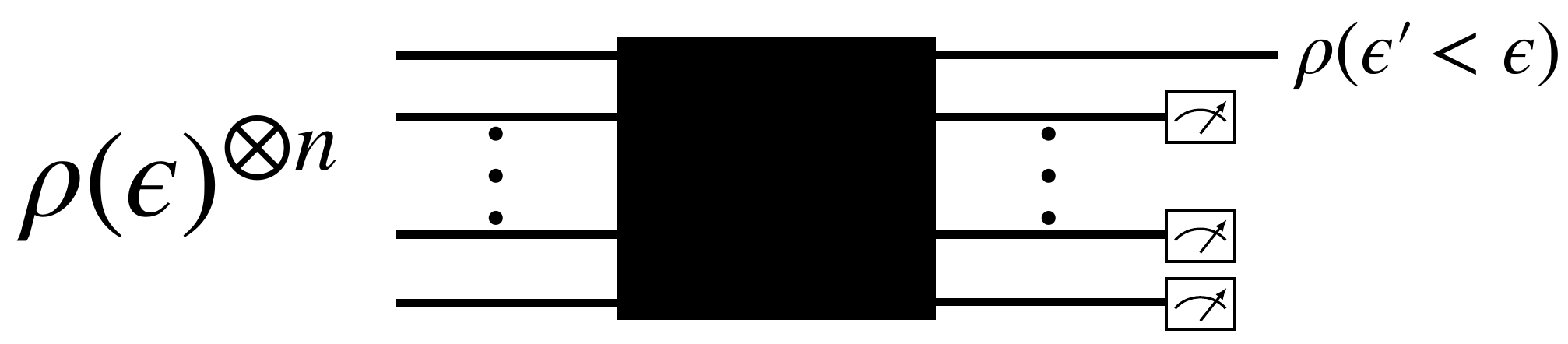}
    \caption{\textbf{Cartoon schematic of distillation scheme.} $n$ copies of a noisy photon state with error rate $\epsilon$ (Eq.~\eqref{eq:single-photon-mixed}), incident upon $n$ spatial rails,  are used to distill a single photon of lower error $\epsilon'< \epsilon$. This is achieved upon post-selection of a particular detection pattern of $n-1$ photons in the measured rails. The black box is at this point unspecified but will be an array of beamsplitters between the rails to enact interference.}
    \label{fig:distillation_cartoon}
\end{figure}

Similarly, for specific applications of LOQC, such as Boson sampling \cite{boson-sampling}, multi-photon interference is the key ingredient to generate a computationally intractable distribution, which is reduced in complexity with distinguishability \cite{boson-sampling-distinguishable}.

It is therefore necessary to be able to generate photons with as high an overlap as possible. In this Letter, we present a technique inspired by magic state distillation \cite{MSD}, which is used to `distill' indistinguishable photons from a photon source which outputs photons that are partly distinguishable {(or in other words, a source with non-unit purity)}.
This task can be phrased in a few equivalent ways, and is related to state purification \cite{opt_purification, experimental_purification} and  discrimination \cite{dist-discrim}.

{Commonly narrowband filters are used to generate heralded highly pure photons from pair sources, however in practice the photon yield becomes prohibitively small at high enough target purity \cite{hom_heralded}. {Moreover, naive filtering of a single photon source, whilst yielding highly pure photons, will be unheralded.} Our scheme instead works under a different paradigm, where independent single photons that are partly distinguishable are used to produce a source of heralded and pure photons, utilizing multi-photon interference.}

A cartoon example of our general idea is shown in Fig.~\ref{fig:distillation_cartoon}, whereby $n$ copies of a noisy photon state are used to produce single photons, with a lower degree of distinguishability. Input photons to the circuit populate spatial modes (horizontal lines), which we will often refer to as `rails', and can be implemented physically via optical fibres, for example. The black box is  a circuit composed of beamsplitters (and possibly other linear optical components), and the output photon is conditioned on the post-selection of a particular measurement outcome, i.e. the detection of $n-1$ photons, in some configuration.
The key observation behind our scheme is that identical photons interfere in a fundamentally different manner than partly distinguishable ones, which can be exploited using beamsplitters, and ultimately used to reduce the distinguishability of noisy photon sources.
{The scheme works so long as the initial purity is above around 60\%.}

\vspace{0.3cm}
\textit{Related Work--} Whilst preparing this manuscript, we became aware of a morally similar scheme proposed by Sparrow and Birchall (SB) in Ref.~\cite{sparrow-thesis}, under the name `HOM filtering'. 
In this scheme, $n\ge 2$ photons are incident upon $n$ rails, which are post-selected upon bunching in a single rail. Photon subtraction is then used to output a single photon of a higher fidelity.
This scheme is conceptually elegant, and results in asymptotic scaling of the error $\epsilon \rightarrow \epsilon/n$. 
However, it is apparent that the scheme becomes prohibitive for even modest $n$, as the probability to measure the desired outcome falls worse than exponentially in $n$\fn{Numerically we find Eq.~\eqref{eq:SB_success} can be approximated by $\exp(-0.29 n^{1.96})$, see Fig.~\ref{fig:sb_ps}}; we compute in App.~\ref{sect:SB_success} the post-selection success probability to be asymptotically (i.e. at error approaching zero)
\begin{equation}
    \label{eq:SB_success}
    P_{p.s.}^{(SB)} \le \frac{n}{2^n} \prod_{m = 2}^n \frac{m}{2^m} = \frac{n^2 (n-1)!}{\sqrt{2^{n^2 + 3n - 2} }},
\end{equation}
meaning huge numbers of photons are required to distill a single purer one (for $n=2,3,4,5,6$, one requires on average 8, 42, 341, 4369, 93206 photons respectively).

Our scheme overcomes two issues identified by SB in their protocol, namely we achieve higher success probabilities (and therefore use fewer photons), and do not require explicit multiple photon subtraction\citedfn{bleed-note}.
Eventually we believe a hybrid scheme can be invoked, as in regimes of higher error, the SB scheme can outperform the present approach, whereas at lower errors, our scheme is most efficient. We will discuss this in the \hyperlink{sect:results}{Results} section.

\vspace{0.3cm}
\hypertarget{sect:theory}{\textit{Theory--}}
An arbitrary single photon state can be written as a sum over modes \cite{photon_modes, mode_structure}:
\begin{equation}
    |\psi\rangle = \sum_{s\in \{h,v\}}\int d\omega\, c_{s,\omega} |s,\omega\rangle = \sum_{i=0}^\infty c_i \hat{a}_i^\dag|\mathbf{0}\rangle=\sum_{i=0}^\infty c_i |\psi_i\rangle.
    \label{eq:single-photon}
\end{equation}
The term after the first equals sign represents the explicit representation over the polarization ($s$ being e.g. horizontal $h$ or vertical $v$) and frequency ($\omega$) domains, and going to the second equals sign we have picked a countable orthonormal basis in the separable Hilbert space to represent the continuous degrees of freedom (and absorbed the $s$ index into the new sum). 
The state $|\mathbf{0}\rangle$ is the vacuum state, and $\hat{a}^\dag_i$ creates a photon in the $i$'th mode, where for now we use the explicit state representation $\hat{a}_i^\dag |\mathbf{0}\rangle=|\psi_i\rangle$. By construction, these basis states are orthogonal $\langle \psi_i|\psi_j\rangle = \delta_{ij}$, and the amplitudes $c_i \in \mathbb{C}$ square sum to 1: $\sum_i |c_i|^2=1$.

We now describe the model of a noisy photon source which is used in this work. A non-ideal photon source will output photons according to Eq.~\eqref{eq:single-photon}, but with realization dependent coefficients $c_i$ (that is, they are different for each generated photon). Without loss of generality we can pick the basis so that the desired mode to populate is the 0'th one, i.e. $|\psi_0\rangle$ is the state which would be generated each time by a perfect photon source. We consider fluctuations around this ideal by assuming the source can generate photons in the 0'th mode with probability $1-\epsilon$, i.e. $\langle |c_0|^2 \rangle = 1-\epsilon$, where the angle brackets indicate the realization average. We will similarly define $p_i := \langle |c_i|^2\rangle$, where $\sum_{i>0} p_i = \epsilon$.
We further make a random phase approximation so that $\langle c_i c_j^*\rangle = 0$ for $i\neq j$, which means the photon source can be equivalently described as a dephased mixture:
\begin{equation}
    \rho(\epsilon) = (1-\epsilon)\, |\psi_0 \rangle \langle \psi_0 | + \sum_{i>0} p_i\, |\psi_i \rangle \langle \psi_i|. 
    \label{eq:single-photon-mixed}
\end{equation}
This approximation amounts to the `error amplitudes' $c_{j>0}=|c_j| e^{i\phi_j}$ receiving a random phase $\phi_j$ (independent of the norm) on each realization.
With this, we can therefore interpret the photon source as generating a photon in the ideal state $|\psi_0\rangle$ with probability $1-\epsilon$, or with probability $\epsilon$ an orthogonal `error mode' is populated (i.e. from one of the $\hat{a}_{i>0}^\dag$). We will similarly call the $|\psi_{i>0}\rangle$ as an `error state' (orthogonal to $|\psi_0\rangle)$.

{We define the indistinguishability within our model as the mean overlap of pure states generated by the source, i.e. $\mathcal{I}:=\mathrm{mean}(|\langle \phi | \psi\rangle|)$. Under our assumptions, this is equivalent to sampling pure states from $\rho$, from which it is easy to show $\mathcal{I}=\mathrm{tr}(\rho^2)$, i.e. it is the purity. The aim of this work is to maximise the {indistinguishability} by minimizing $\epsilon$.}

To simplify the analysis, we can consider the small error (small $\epsilon$) limit. At sufficiently small $\epsilon$ it is unlikely to observe more than one error state according to the above statistical description; if we draw $n$ samples from distribution $\rho$\citedfn{sources}, we either get $n$ copies of $|\psi_0\rangle$, or $n-1$ copies of $|\psi_0\rangle$, and one copy of some orthogonal error state $|\psi^\perp\rangle$ (i.e. $|\psi^\perp\rangle$ is one of the $|\psi_{i>0}\rangle$). Note, in our subsequent analysis we will still take into account the cases when more than one error mode is populated, but for now we can work in the limit of only single errors, for convenience.
We can write the $n$ photon state, to first order as {(see App.~\ref{sect:rho-deriv})}
\begin{equation}
    \rho^{\otimes n} = (1-\epsilon)^n |\Psi_0\rangle \langle \Psi_0| + \epsilon (1-\epsilon)^{n-1} \sum_{k=1}^n |\Psi_k\rangle \langle \Psi_k | + O(\epsilon^2),
    \label{eq:multi-photon-mixed}
\end{equation}
where we have introduced notation $|\Psi_0\rangle = |\psi_0\rangle^{\otimes n}$ and $|\Psi_k\rangle = |\psi_0\rangle^{\otimes (k-1)} |\psi^\perp \rangle |\psi_0\rangle^{\otimes (n-k)}$. {The error term $O(\epsilon^2)$ contains the states of $n$ photons composed of $n-2$ copies of $|\psi_0\rangle$, and two error states $|\psi_{i>0}\rangle$.}
The tensor structure comes from the spatial mode representation, as in Fig.~\ref{fig:distillation_cartoon}. For now we write the error state generically as $|\psi^\perp\rangle$, as we will later see at first order it is unimportant for our analysis which particular error mode $i>0$ is populated in state $|\Psi_k\rangle$.

In order to enact interference between photons of the above form, we will utilise a beamsplitter. In our notation a beamsplitter is described by 4 parameters, and acts on (spatial) mode creation operators $\hat{a}^\dag, \hat{b}^\dag$ as follows:
\begin{equation}
\begin{split}
    & \hat{a}^\dag \rightarrow e^{i(\phi_0 + \phi_R)}\sin(\theta) \hat{a}^\dag + e^{i(\phi_0 + \phi_T)}\cos (\theta) \hat{b}^\dag \\
    & \hat{b}^\dag \rightarrow e^{i(\phi_0 - \phi_T)}\cos (\theta) \hat{a}^\dag - e^{i(\phi_0 - \phi_R)}\sin(\theta) \hat{b}^\dag.
\end{split}
\label{eq:beamsplitter}
\end{equation}
 We assume the parameters $\{\theta, \phi_{0,R,T}\}$ are agnostic to the impinging photons internal state \cite{agnostic-beamsplitter}, and therefore any single photon incident upon such a beamsplitter will be ‘split’ in the same manner as any other.
 A 50:50 beamsplitter refers to the case $\theta=\pi/4$, where there is equal transmission to the other mode ($T$), or reflection to the same mode ($R$).
Throughout we use the the convention for the phases $\phi_0=\pi/2, \phi_R=-\pi/2, \phi_T=0$.

Since we utilise optical components that are state-agnostic, and any single photon in state $|\psi_{i>0}\rangle$ will not interfere with the ideal state $|\psi_0\rangle$ (by orthogonality), it has no bearing on the output statistics of a circuit of form Fig.~\ref{fig:distillation_cartoon}
which particular error mode $i>0$ is actually populated when state $|\Psi_k\rangle$ is sampled from $\rho^{\otimes n}$. For this reason we can write the single error state simply as $|\psi^\perp\rangle$, as mentioned above.

Now that we have described the basic components in our construction, all that remains is to outline the post-selection over detection events.
We will require access to photon number resolving detectors which we assume are ideal; it will always detect the exact number of photons present (though it will in fact be enough to distinguish between 0,1,2,3 photons, which will be clear later).
The post-selection on a detection event of $m$ photons can be described by taking the partial trace of the measured rail(s) after applying a measurement operator on the state \cite{sparrow-thesis, povm}.
If before measurement the state is $\rho$, and we place a detector at the $k$'th rail to detect $m$ photons, the post-selected state will be $\mathrm{Tr}_k[\Pi_k^{(m)} \rho \Pi_k^{(m)}] / N$,
where $\Pi_k^{(m)}$ sums over all rank 1 projectors onto pure states which contain $m$ photons in the $k$'th rail. $N$ is for normalization.

\vspace{0.3cm}
\hypertarget{sect:results}{\textit{Results--}} The central question we wish to answer is whether one can engineer the schematic diagram Fig.~\ref{fig:distillation_cartoon} with a suitable number $n$ of photons, and linear optical components in the black box, so that the output state has less error than Eq.~\eqref{eq:single-photon-mixed}, upon a suitable post-selection. If one can do this, the process can be repeated indefinitely until arbitrary accuracy (i.e. $\epsilon$ is arbitrarily small). 

From our studies, this in fact defines a large class of optical circuit of varying numbers of photons and linear optical components. We however will focus our attention on the `best' performing that we found (where here best has a precise meaning, in terms of the number of photons required to distill a photon to some particular accuracy). Indeed, there is scope for the discovery of improved circuits. 
We will assume all components and detectors are perfect, so that the only source of  error is in the photon generator, but discuss such errors in App.~\ref{sect:errors}.

\begin{figure}
    \centering
    \includegraphics[width=0.98\columnwidth]{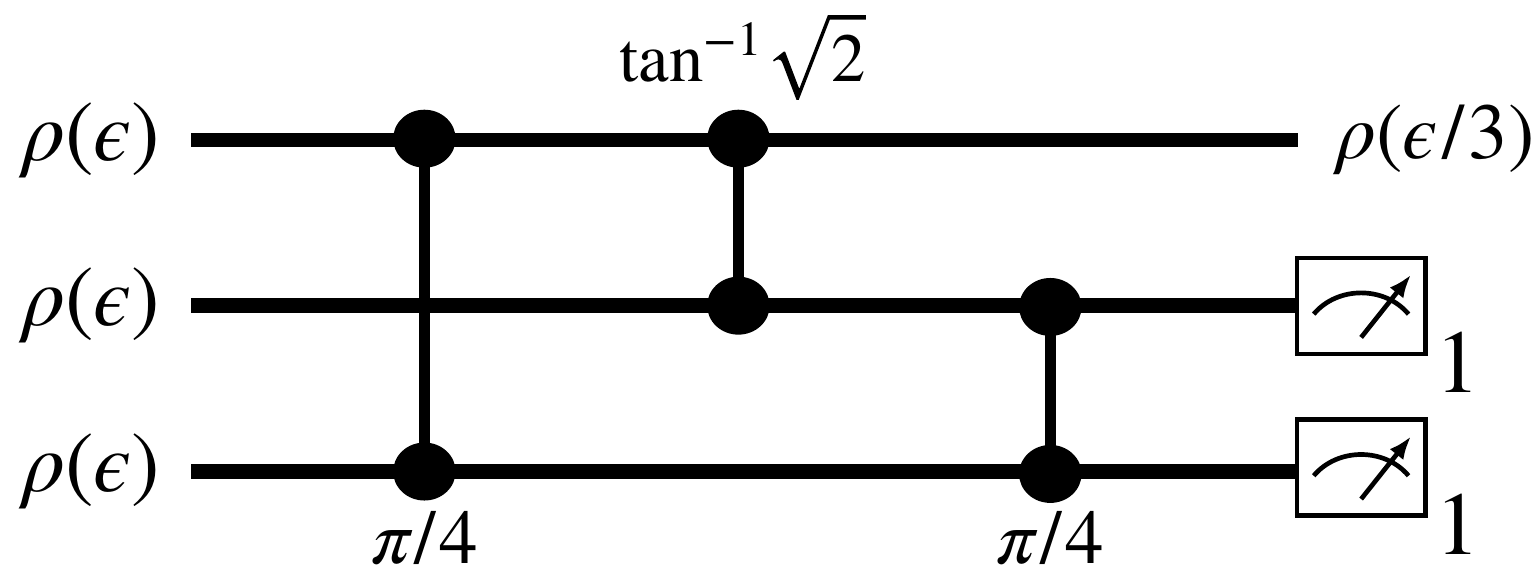}
    \caption{\textbf{Three photon distillation scheme.} A successful measurement corresponds to a single photon registered in each of the two measured rails (indicated by the `1' subscript on the detectors).
    The vertical lines with black circles represent beamsplitters between the rails on which the black circles intersect.
    The first and third beamsplitters are 50:50 ($\pi/4$ in the diagram), and the middle is asymmetric with $\theta = \tan^{-1}\sqrt{2}\approx 0.955$ (less likely to transmit). In the asymptotic limit of small $\epsilon$, the error is reduced by a factor of $1/3$, and post-selection succeeds with probability $1/3$.
    }
    \label{fig:n_3_circ}
\end{figure}

The circuit of present interest is shown in Fig.~\ref{fig:n_3_circ}, composed of three rails (each taking one incident photon), and three beamsplitters, two which are symmetric, and one which is asymmetric, biased to higher reflectivity (to stay in same mode). Note permutations of this circuit also perform identically (keeping the angle of the middle beamsplitter $\tan^{-1}\sqrt{2}$).

First let us consider the ideal input of three identical photons in state $|\psi_0\rangle$ sampled from $\rho$, which we will denote using occupation number (Fock) representation over the rails as $|1,1,1\rangle$. This input occurs with probability $(1-\epsilon)^3$. The output of the circuit, before measurement is (up to a global phase)
\begin{equation}
\label{eq:output_state}
    \frac{1}{\sqrt{3}}|1,1,1\rangle - \frac{\sqrt{2}}{3}(i|3,0,0\rangle - |0,3,0\rangle + i|0,0,3\rangle),
\end{equation}
which has probability of $1/3$ to obtain the correct post-selected state\fn{Note this scheme can also be used to prepare the superposition $|3,0\rangle + |0,3\rangle$, upon measuring zero photons in the middle rail of Eq.~\eqref{eq:output_state}, which occurs with probability 4/9, which can be considered a 3 photon generalization of the standard HOM state.}.

If on the other hand a single error state is present, i.e. one of $\{|\Psi_k\rangle \}_{k=1}^3$ is sampled (each occurring with probability $\epsilon (1-\epsilon)^2$), the output in the relevant subspace before measurement, is $\frac{1}{\sqrt{27}}\sum_{k=1}^3|\Psi_k\rangle$, up to a phase.
The post-selection therefore succeeds with total probability $1/9$, and the outputted (unmeasured) photon is ideal $|\psi_0\rangle$ with post-selected probability $2/3$ {(see App.~\ref{sect:eq_post_sel} for more information)}.

The key observation behind the scheme is that the ideal input is successfully post-selected upon three times as often than the case where an error is present ($1/3$ {vs} $1/9$), which allows the errors to be filtered out, approximately at a rate of 1/3 error reduction per round.

One can produce an upper bound on the error reduction (see App.~\ref{sect:eq_post_sel}), $\epsilon \rightarrow \epsilon'$ under the scheme:
\begin{equation}
\label{eq:post-sel}
    \epsilon' \le \frac{\epsilon}{3} \frac{1 + 2\epsilon}{1 -2\epsilon + 3\epsilon^2 - \epsilon^3} =  \frac{\epsilon}{3} + \frac{4\epsilon^2}{3} + O(\epsilon^3).
\end{equation}
The reason this is a bound, instead of equality, is that  the error reduction depends on the specifics of the distribution of errors in  Eq.~\eqref{eq:single-photon-mixed}. In App.~\ref{sect:eq_post_sel} we also produce a lower bound on the error, $\epsilon' \ge \frac{\epsilon}{3} + \frac{2\epsilon^2}{3} + O(\epsilon^3)$.
The scheme can be used to reduce errors ($\epsilon' < \epsilon)$ so long as the initial error $\epsilon$ is below around 43\%.

\begin{figure}
    \centering
    \includegraphics[width=0.98\columnwidth]{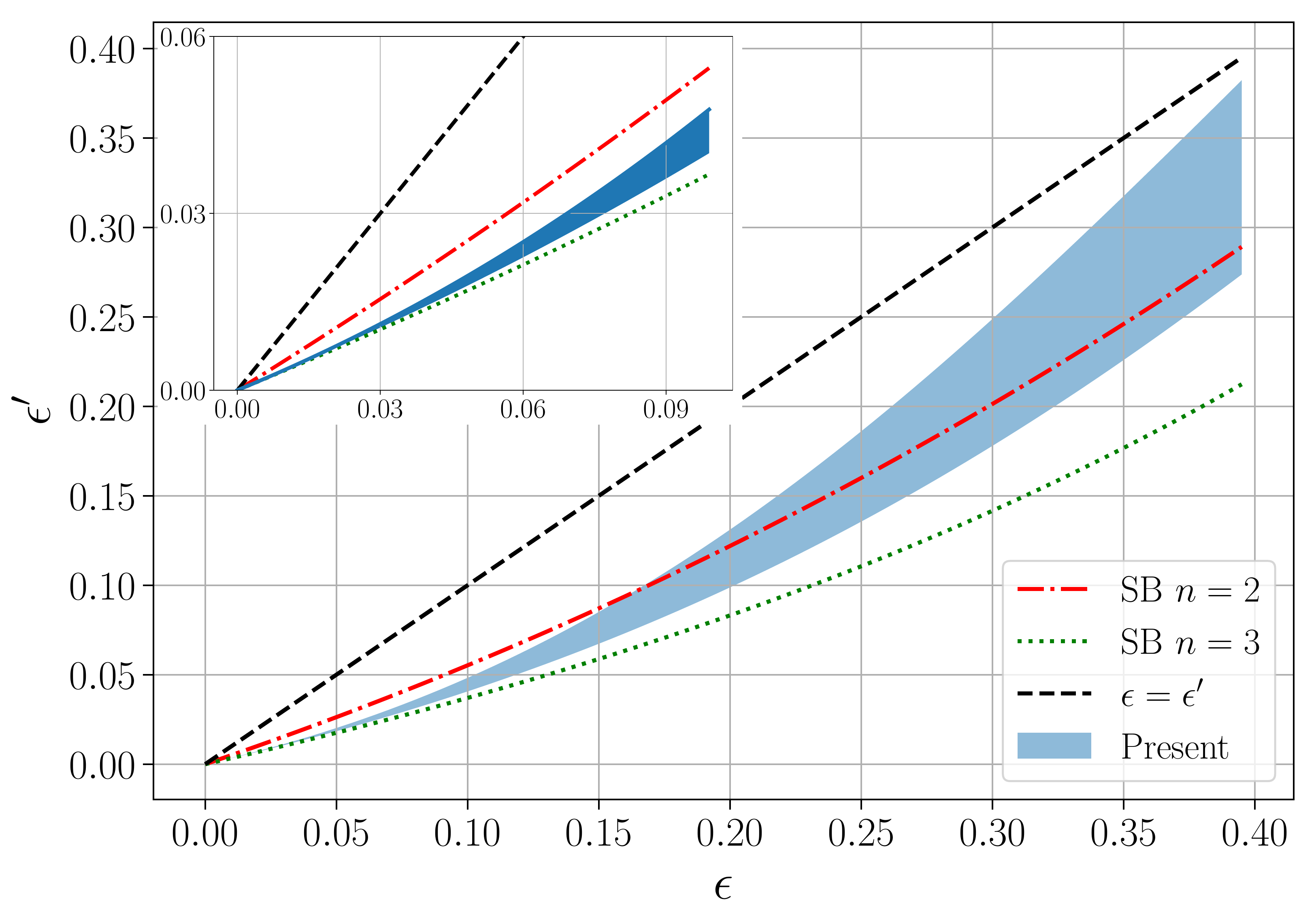}
    \caption{\textbf{Error reduction comparison of our scheme (`Present'), and those of SB for $n=2,3$}. Given a photon source with error $\epsilon$ (Eq.~\eqref{eq:single-photon-mixed}), the post-selected output has error $\epsilon^\prime$ {(non-asymptotic Eq.~\eqref{eq:post-sel} used for the upper-bound)}. The shaded region indicates the upper and lower bound on the error reduction of our scheme, as discussed in the main text (and App.~\ref{sect:eq_post_sel}).
    For SB, we use the best case error reduction, Eq.~(7.11) in Ref.~\cite{sparrow-thesis} (also see Eq.~\ref{eq:SB_error}). Inset: Zoom in on region $\epsilon < 0.1$. 
    }
    \label{fig:error_reduction}
\end{figure}

The error reduction capabilities of our scheme is shown in Fig.~\ref{fig:error_reduction}, where we also compare to the SB protocol for $n=2$ {which as we will see is the most efficient SB protocol}, and $n=3$ ({same number of photons per round as the present approach}). We see our scheme outperforms SB for $n=2$ for errors less than around 15\%, and that our scheme converges with SB $n=3$ at around 5\% error. Note, for the SB scheme we plot the best case error reduction, whereas in reality it may perform worse than this, depending on the distribution of error modes, see Ref.~\cite{sparrow-thesis} (though for small $\epsilon$ the difference becomes negligible).

In App.~\ref{sect:eq_post_sel} we compute the probability of obtaining a valid post-selection measurement outcome (i.e. detection of a single photon at each of the two detectors), which scales as $(1-2\epsilon)/3 + O(\epsilon)^2$. {Fig.~\ref{fig:post_sel} compares this to the SB $n=2,3$ protocols which have a lower post-selection probability, leading to a greater resource requirement}. 
Since our scheme consumes 3 photons per use, we require around 9 photons to distill a single purer one to $1/3$ the error. In comparison to SB for $n=2,3$, around 8 and 42 photons are required respectively to obtain $1/2, 1/3$ error respectively.
In the asymptotic error limit (which practically is for $\epsilon \lesssim 0.05$), one can compute the number of photons required to distill a photon to target error $\epsilon'$ as $O((\frac{\epsilon}{\epsilon'})^2)$\fn{The calculation of $O((\epsilon/\epsilon')^2)$ photons comes simply by noting each iteration of the scheme requires asymptotically 9 photons, to reduce the error by 1/3. If we wish to obtain error $\epsilon'$, we require $r$ iterations where $\epsilon' = \epsilon/3^r$, which consumes $9^r = (\epsilon/\epsilon')^2$ photons. The `big $O$' notation captures the constant overhead when $\epsilon/\epsilon'$ is not an exact power of 3.}. In comparison to SB $n=2,3,4$, the exponent is $3, 3.4, 4.2$ respectively.
This implies in the asymptotic limit our scheme is the most efficient.

Lastly, we wish to mention we also discovered an $n=4$ photon circuit (see App.~\ref{sect:n_4_circ}), which is essentially a generalization of the presented $n=3$ circuit (though with only 50:50 beamsplitters), which can reduce errors by $\epsilon/4$, at the expense of a lower success probability -- asymptotically 1/4 -- meaning around 16 photons are required on each iteration, and still $O((\frac{\epsilon}{\epsilon'})^2)$ photons to distill to error $\epsilon'$. 

\vspace{0.3cm}
\textit{Discussion--} 
{We briefly comment here that the scheme has some attractive properties for experimental implementation, which is discussed in more detail in App.~\ref{sect:errors}. 
In particular, there is a natural robustness to detection errors, as well as control errors. We also mention the protocol can also be trivially implemented in the case where the individual photons come from different physical sources\citedfn{sources}. For example, single photons of modest purity and reasonably high production rate could be generated from heralded filtered {spontaneous parametric down-conversion (SPDC)} pairs \cite{improving-spdc}, and then boosted to a high target fidelity via distillation, which crucially, are still heralded.}

Overall, in realistic scenarios, various errors will limit the upper bound on the indistinguishability that can be reached by our scheme, and a natural follow up can investigate robustness to these in practical settings.
Additionally, the techniques presented here, we believe, have a diverse range of application, and can be utilized directly in resource state generation to construct circuits that are naturally resilient to distinguishability and loss errors, using similar mechanisms.

\vspace{0.3cm}
\textit{Acknowledgements--} We thank Joseph Altepeter, Ryan Bennink, Patrick Birchall, Warren Grice, and Raphael Pooser for helpful discussions on the scheme presented, and for providing references to related literature.

We are also extremely grateful to fellow QuAIL team members Eleanor G. Rieffel, Shon Grabbe, Zhihui Wang and Salvatore Mandr\'{a} for support and various discussions on this work.

We are thankful for support from NASA Academic Mission Services, Contract No. NNA16BD14C.

We acknowledge support from DARPA, under DARPA-NASA IAA 8839, annex 129.

\bibliography{refs.bib}

\begin{thebibliography}{33}
\bibitem [{ble()}]{bleed-note}%
  \BibitemOpen
  \href@noop {} {}\bibinfo {note} {{In principle photon bleeding can be invoked
  to achieve success probability approaching 1 for the photon subtraction phase
  \cite{bleeding} in SB, at the cost of additional circuit depth. The
  feasibility of this however depends on loss and detection errors which could
  accumulate.}}\BibitemShut {Stop}%
\end{thebibliography}

\appendix

\section{Derivation of main text Eq.~(1)\label{sect:SB_success}}

We first direct the reader to Ch.~7.2 of Ref.~\cite{sparrow-thesis} for explicit outline of the SB protocol.

We consider the case where all photons are identical (i.e. case of zero distinguishability), which is the dominant contribution to the post-selection probability as the error $\epsilon \rightarrow 0$.
There are two independent calculations to arrive at Eq.~(1) of the main text.

The first involves computing the probability for an input of $m$ and $1$ photons in separate rails to bunch, that is, $|m,1\rangle \rightarrow |m+1,0\rangle$, upon post-selection by measuring 0 photons in the second rail.

To do that, let us denote the initial state using creation operators as
\begin{equation*}
    |m, 1\rangle =: \frac{1}{\sqrt{m!}} (\hat{a}^\dag)^m\hat{b}^\dag|{0}, {0}\rangle .
\end{equation*}

Under our convention outlined in the main text, a 50:50 beamsplitter acts as (though the choice of phase angles is unimportant here)
\begin{equation*}
\begin{split}
    & \hat{a}^\dag \rightarrow \frac{1}{\sqrt{2}}(\hat{a}^\dag + i \hat{b}^\dag) \\
    & \hat{b}^\dag \rightarrow \frac{1}{\sqrt{2}}( i\hat{a}^\dag +  \hat{b}^\dag).
\end{split}
\end{equation*}

We wish to compute the amplitude of the coefficient $(\hat{a}^\dag)^{m+1}$ in the expansion
$(\hat{a}^\dag+i\hat{b}^\dag)^m(i \hat{a}^\dag+\hat{b}^\dag)/\sqrt{2^{m+1}}$, which one can see by inspection is simply $i/\sqrt{2^{m+1}}$.
The \textit{states} amplitude is computed via 
\begin{equation*}
    \frac{1}{\sqrt{m!}}\frac{i}{\sqrt{2^{m+1}}}(\hat{a}^\dag)^{m+1}|{0},{0}\rangle = i\sqrt{\frac{m+1}{2^{m+1}}}|m+1,0\rangle,
\end{equation*}
and therefore sampled with probability $p_m = \frac{m+1}{2^{m+1}}$.

For the $n$ photon circuit, the above is applied iteratively, starting with $m=1$, up to $m=n-1$, which gives the total bunching probability
\begin{equation*}
    \prod_{m=1}^{n-1}p_m = \prod_{m=2}^n\frac{m}{2^m}.
\end{equation*}

The other independent contribution comes from the photon subtraction after the above post-selection scheme, taking $|n, 0\rangle$ to $ |1, n-1\rangle$ (again by measuring the second rail, this time to find $n-1$ photons). By similar analysis, this can be shown to give probability $n/2^n$, which combining with the above result, completes the proof of Eq.~(1). Note however, as also mentioned in Footnote\citedfn{bleed-note}, this photon subtraction stage can be made almost deterministic at the expense of greater circuit depth, by bleeding. This works by `weakly' subtracting photons one at a time (using beamsplitters with very small transmisivity) until the desired number are siphoned off, as explained in Ref.~\cite{bleeding}. Since one requires the use of additional beamsplitters, one would need to weigh up the benefits when losses, dark counts and other errors are present.

Note that the case of partial distinguishability ($\epsilon > 0$) will result in a lower success probability, hence why we write main text Eq.~(1) as an upper bound.

\begin{figure}
    \centering
    \includegraphics[width=0.98\columnwidth]{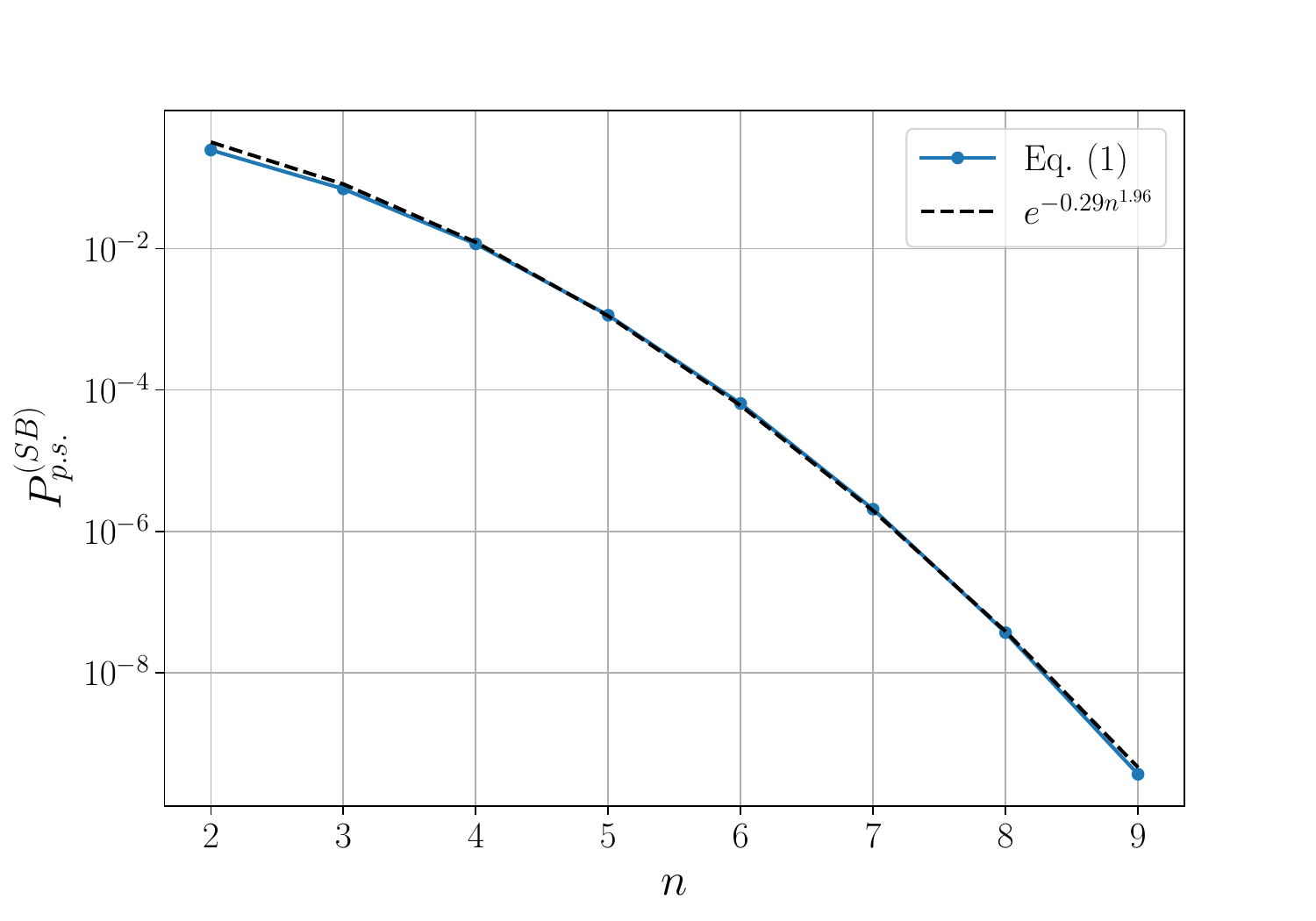}
    \caption{\textbf{Asymptotic SB post-selection probability.} Also plotted is a numerical approximation which decreases exponentially in $n^{1.96}$.}
    \label{fig:sb_ps}
\end{figure}

We plot Eq.~(1) in Fig.~\ref{fig:sb_ps}, as well as a numerically found approximation shown in the legend, which decreases exponentially in $n^{1.96}$.

\section{{Derivation of main text Eqs.~(3,4)} \label{sect:rho-deriv}}
We begin with Eq.~(2) of the main text, that a single photon state is of the form $|\psi\rangle = \sum_{i=0}^\infty c_i |\psi_i\rangle$, where the $|\psi_i\rangle$ are orthonormal. Under our assumptions, the $c_i$ are randomly distributed according to $\langle |c_i|^2\rangle =: p_i$ (and $p_0=1-\epsilon$), where the angle brackets denote a realization average.
We also assume $\langle c_i c_j^*\rangle = 0$, which is a random phase approximation. 
As we will see, this is equivalent to the `random source model'  \cite{sparrow-thesis}. We also mention, this could be the average description of a single photon source, or indeed multiple sources, which holds as long as the above properties are satisfied.

We can arrive at Eq.~(3) by first considering the density matrix of a single realization
\begin{equation*}
    |\psi\rangle \langle \psi|_k = \sum_i |c_i^{(k)}|^2 |\psi_i\rangle \langle \psi_i| + \sum_{i,j} c_i^{(k)} c_j^{(k) *} |\psi_i\rangle \langle \psi_j|,
\end{equation*}
where $k$ indicates this is the $k$'th realization. We now average over the $k$, to compute the density matrix $\rho  = \lim_{N\rightarrow \infty} \frac{1}{N} \sum_{k=1}^N |\psi\rangle \langle \psi|_k = \langle |\psi\rangle \langle \psi|_k  \rangle_k$, which leads directly to Eq.~(3). 

Now we consider taking multiple tensor copies of $\rho(\epsilon)$ of Eq.~(3). Let's re-write this as $\rho = (1-\epsilon)\rho_0 + \epsilon \rho^\perp$, where $\rho_0 := |\psi_0\rangle \langle \psi_0|$ and $\rho^\perp = \frac{1}{\epsilon} \sum_{i>0} p_i |\psi_i\rangle \langle \psi_i|$ are both normalized density matrices.

Then
\begin{equation*}
\begin{split}
    & \rho^{\otimes n} = ((1-\epsilon)\rho_0 + \epsilon \rho^\perp)^{\otimes n} =  (1-\epsilon)^n \rho_0^{\otimes n} +\\  
    & \sum_{k=1}^n
    (1-\epsilon)^{k-1} \rho_0^{\otimes (k-1)} \otimes \epsilon \rho^\perp \otimes (1-\epsilon)^{(n-k)} \rho_0^{\otimes (n-k)} \\
    & + \dots
\end{split}
\end{equation*}
where the trailing dots indicate tensor terms containing two or more copies of ${\rho}^\perp$, i.e. terms on the order $O(\epsilon^2)$.

Now, when considering pure state sampling from the distribution  $\rho^{\otimes n}$ to order $\epsilon$, we see we either get $n$ copies of $|\psi_0\rangle$ with probability $(1-\epsilon)^n$, or $n-1$ copies of $|\psi_0\rangle$, and one copy of a $|\psi_{i>0}\rangle$, which occurs with probability $(1-\epsilon)^{n-1} p_i$. Since $\sum_i p_i=\epsilon$, with probability $\epsilon (1-\epsilon)^{n-1}$ we sample exactly one `error state', $|\psi_{i>0}\rangle$.
However, since all of the $|\psi_{i>0}\rangle$ are orthogonal to $|\psi_0\rangle$, the output of a linear optical circuit does not depend on which particular $|\psi_{i>0}\rangle$ was sampled; the output statistics do not depend on the choice of $i$.
This allows us to a adopt a more meaningful notation, that with probability $\epsilon (1-\epsilon)^{n-1}$, \textit{some} particular error state is sampled, call it $|\psi^\perp\rangle$. This yields exactly Eq.~(4) of the main text.

This first order approximation is convenient for understanding how the protocol works, and this interpretation becomes more accurate at `small' $\epsilon$. However, we stress that the error bounds we derive do not make any assumptions on the smallness of $\epsilon$.

\section{Derivation of main text Eq.~(7) \label{sect:eq_post_sel}}

 First note that the transfer matrix for the circuit Fig.~2 of the main text is
\begin{equation}
    \left(\begin{array}{ccc}
    1/\sqrt{3} & i/\sqrt{3} & i/\sqrt{3}  \\
    -1/2 + i/\sqrt{12}& 1/\sqrt{3} & -1/\sqrt{12}+i/2 \\
    -1/\sqrt{12} + i/2 & i/\sqrt{3} & 1/2 - i/\sqrt{12}
\end{array} \right),
\label{eq:n_3_tm}
\end{equation}
from which all output quantities can be computed.

Now, recall that our error model is derived from main text Eq.~(3), which is equivalent to receiving the ideal photon state $|\psi_0\rangle$ with probability $1-\epsilon$, or an orthogonal error state $|\psi^\perp\rangle$ with probability $\epsilon$.

There are four cases to consider to derive a bound on the output post-selected fidelity from our circuit Fig.~2, which we will denote as 
\begin{equation}
    f' := \langle \psi_0 | \rho_{p.s.} |\psi_0\rangle =: 1-\epsilon',
\end{equation}
where $\rho_{p.s.}$ is the post-selected density matrix.
Since the output fidelity will depend on the error model, namely the distribution $\{p_i\}$ in Eq.~(3), we will find lower and upper bounds on the fidelity.

\textbf{0 errors:} With probability $(1-\epsilon)^3$ no error modes are populated, and the input to the circuit is simply $|\psi_0\rangle^{\otimes 3}$. Using Eq.~(6) we see the post-selection succeeds with probability $1/3$, of which the outputted photon is the ideal one.

\textbf{1 error:} {With probability $3\epsilon(1-\epsilon)^2$ a single error mode is populated, and the input to the circuit is one of the $\{|\Psi_k\rangle\}_{k=1}^3$, where
\begin{equation*}
    \begin{split}
        & |\Psi_1\rangle := |\psi^\perp \rangle \otimes |\psi_0\rangle \otimes |\psi_0 \rangle \\
        & |\Psi_2\rangle := |\psi_0\rangle \otimes |\psi^\perp \rangle \otimes |\psi_0\rangle \\
        & |\Psi_3\rangle := |\psi_0\rangle \otimes |\psi_0\rangle \otimes|\psi^\perp \rangle .
    \end{split}
\end{equation*}
The output can be computed using the transfer matrix, Eq.~\eqref{eq:n_3_tm}, by defining the above states using Fock notation: $c_1^\dag(\psi^\perp)c_2^\dag(\psi_0)c_3^\dag(\psi_0)$ (and permutations thereof). Here $c_i^\dag(\psi)$ creates a single photon in state $|\psi\rangle$ in the $i$'th rail.
Expanding this out using the transfer matrix, one finds the output state, in the relevant subspace of a single photon per rail, is $\frac{1}{27}\sum_k |\Psi_k\rangle$. Measurement of the final two rails collapses onto one of the three distinct outcomes $|\Psi_k\rangle$ (as discussed in the main text) with equal probability $1/27$ each, yielding $1/9$ as the total post-selection probability. However, as is clear from the equations above, only when $|\Psi_{2,3}\rangle$ are measured do we output the desired photon state $|\psi_0\rangle$ in rail 1, i.e. upon post-selection, $2/3$ of the output photons are in the desired state, and $1/3$ is a `garbage' orthogonal state.}

\textbf{2 errors:} With probability $3\epsilon^2(1-\epsilon)$ two error modes are populated. Within this there are two sub-cases to consider, i) when the error modes are the same, $|\psi_0, \psi_1, \psi_1\rangle$, or ii) when they are unique $|\psi_0, \psi_1, \psi_2\rangle$ (and permutations thereof). Note we picked the labels $1,2$ for the error modes, however this is completely arbitrary, and we can do this without loss of generality.
We find that the former case yields a valid measurement pattern with probability $1/9$, whereas the latter is $2/9$. In both cases, the outputted photon is the ideal one only $1/3$ of the time.

\textbf{3 errors:} With probability $\epsilon^3$ three error modes are populated. There are technically three classes to consider: modes are all distinct, two the same, or three the same. Of course, in all three cases, there is 0 chance of the ideal photon $|\psi_0\rangle$ being outputted. Since we seek a bound on the output fidelity, we wish to find the case which minimizes/maximises the post-selection probability. When all three error modes are identical the post-selection success probability is maximal for this case, with probability $1/3$ (for the same reason as the 0 error case). The converse case achieves lowest post-selection success with probability $1/9$, occurring when two modes are identical.

The above 4 cases give all possible sampling outcomes, and we can use this to find a lower bound on the output fidelity $f'$, which is simply the ratio of the post-selected probability to obtain the ideal photon state $|\psi_0\rangle$, normalized by the total post-selection success probability. This is achieved taking the $2/9$ post-selection case for two errors, and $1/3$ for three errors, and yields, writing the terms suggestively:
\begin{equation}
\begin{split}
    & f' \ge \frac{ (1-\epsilon)^3 \frac{1}{3} + 3\epsilon(1-\epsilon)^2 \frac{1}{9}\frac{2}{3} + 3\epsilon^2 (1-\epsilon)\frac{2}{9}\frac{1}{3} }{\frac{1}{3}[(1-\epsilon)^3+  \epsilon(1-\epsilon)^2 + 2\epsilon^2(1-\epsilon) + \epsilon^3]} \\
    & = 1 - \frac{\epsilon}{3} - \frac{4\epsilon^2}{3} + O(\epsilon^3).
\end{split}
\end{equation}
This can be rearranged to Eq.~(7) of the main text.

We can similarly obtain an upper (lower) bound on the fidelity (error), with final result
\begin{equation}
    \epsilon' \ge \frac{\epsilon}{3 - 6\epsilon + 6\epsilon^2 -2\epsilon^3} = \frac{\epsilon}{3} + \frac{2\epsilon^2}{3} + O(\epsilon^3).
\end{equation}

\begin{figure}
    \centering
    \includegraphics[width=0.98\columnwidth]{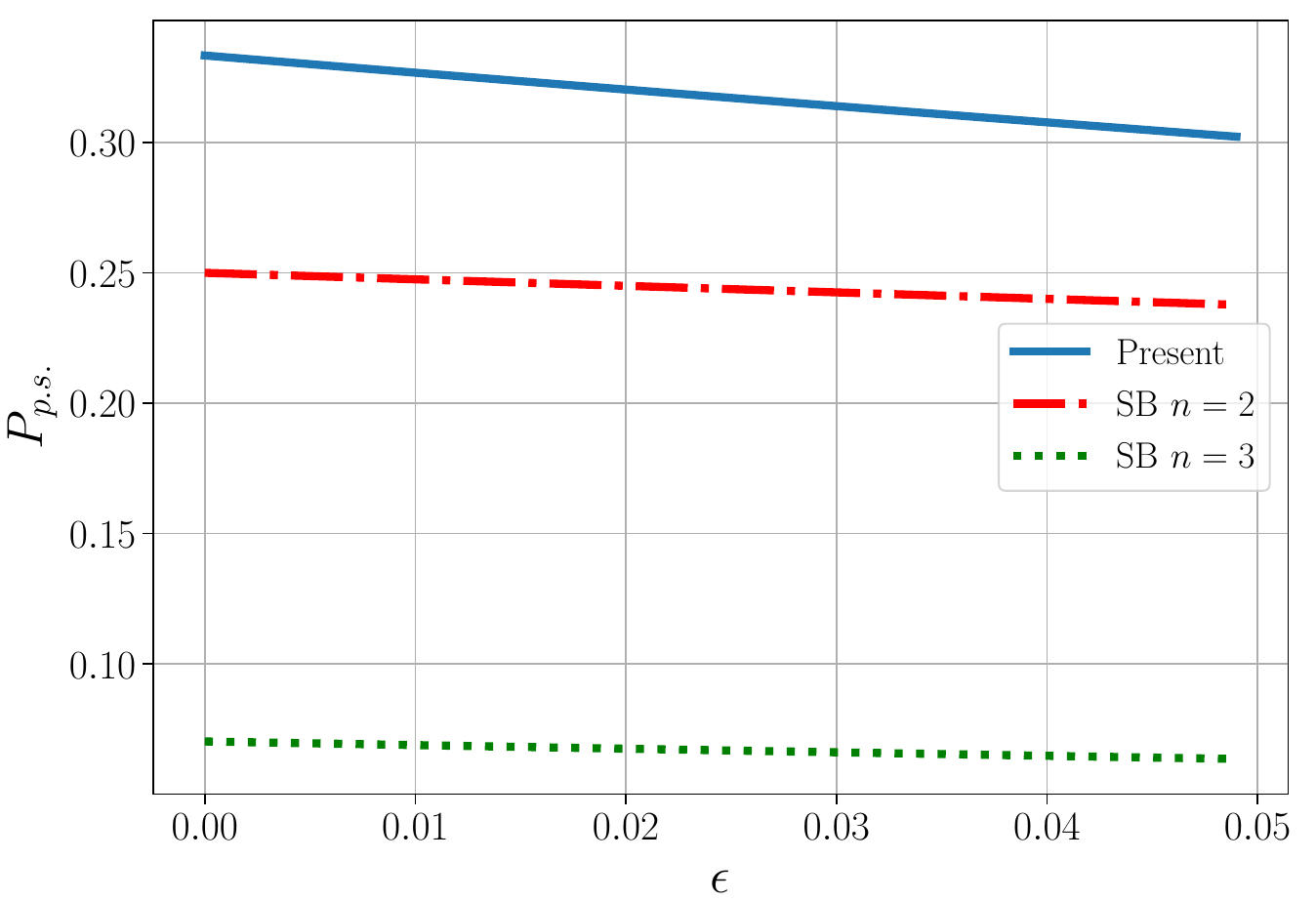}
    \caption{\textbf{Post-selection success probability}. Scaling of the success probability to obtain the desired output as a function of the error. In the low $\epsilon$ limit, our scheme succeeds with probability $1/3$, meaning on average the protocol should be run 3 times, consuming 9 photons, for successful generation of a higher fidelity photon. The $\epsilon=0$ case is that of main text Eq.~(1) for the SB protocols.
    }
    \label{fig:post_sel}
\end{figure}

In the same manner we find a lower bound on the post-selection success probability,
\begin{equation}
\begin{split}
    & P_{p.s.} \ge \frac{1}{3}[(1-\epsilon)^3+  \epsilon(1-\epsilon)^2 + \epsilon^2(1-\epsilon) + \frac{1}{3} \epsilon^3] \\
    & = \frac{1}{3}(1 - 2\epsilon + 2\epsilon^2 - 2\epsilon^3),
\end{split}
\end{equation}
which is shown in Fig.~\ref{fig:post_sel}, also including the result for SB $n=2,3$.

\textit{Calculation of fidelity for SB $n=2,3$:}
Via a similar analysis as above we can find an explicit version of Eq.~(7.11) in Ref.~\cite{sparrow-thesis} for the case $n=2,3$ (which is what we plot in main text Fig.~3).

The result is, writing the error $\epsilon' = 1 - f'$:
\begin{equation}
\begin{split}
        & \epsilon'_{SB, n=2} \ge \frac{\epsilon}{2 - 2\epsilon + \epsilon^2} = \frac{\epsilon}{2} + \frac{\epsilon^2}{2} + O(\epsilon^3), \\
        & \epsilon'_{SB, n=3} \ge  \frac{2\epsilon - 2\epsilon^2 + \epsilon^3}{6 - 12\epsilon + 9\epsilon^2 - 2\epsilon^3}= \frac{\epsilon}{3} + \frac{\epsilon^2}{3} + O(\epsilon^3).
        \label{eq:SB_error}
\end{split}
\end{equation}
As discussed in Ref.~\cite{sparrow-thesis}, this is the best case performance of their protocol (the bound becomes tight for small $\epsilon$).

\section{Four Photon Circuit \label{sect:n_4_circ}}

\begin{figure}
    \centering
    \includegraphics[width=0.98\columnwidth]{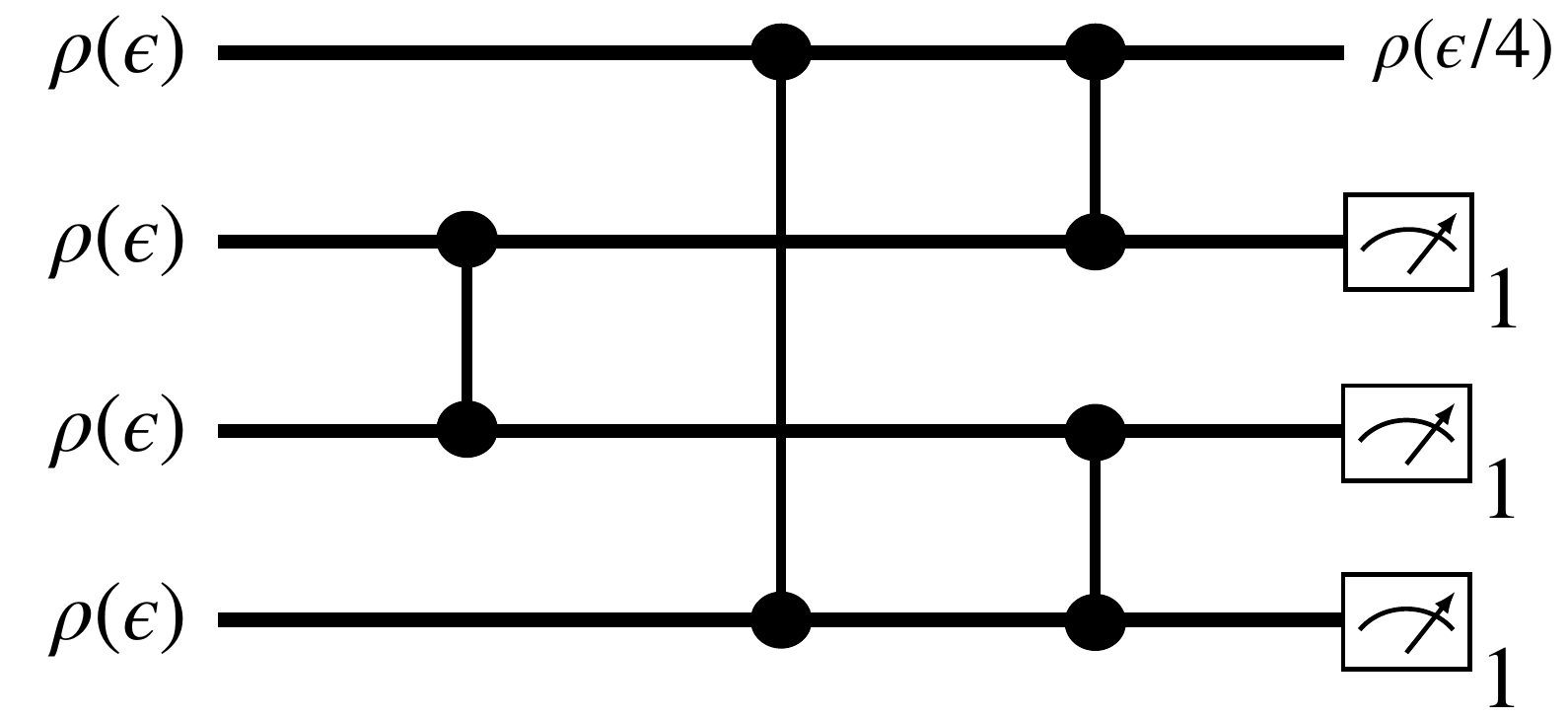}
    \caption{\textbf{Four photon distillation scheme} A successful measurement corresponds to a single photon registered in each of the three measured rails (indicated by the `1' subscript on the detectors).
    Here all beamsplitters are 50:50. In the asymptotic limit of small $\epsilon$, the error is reduced by a factor of $1/4$, and post-selection succeeds with probability $1/4$.}
    \label{fig:n_4_circ}
\end{figure}

Here we outline an equivalent circuit as shown in the main text, for $n=4$, where the error reduction scales as $\epsilon/4$, and post-select success probability $1/4 - 3\epsilon/4 + O(\epsilon^2)$. The circuit uses only 50:50 beamsplitters (in contrast to main text Fig.~2) and is shown in Fig.~\ref{fig:n_4_circ}. This is due to the symmetry of the circuit (the $n=3$ circuit requires the asymmetric beamsplitter to effectively symmetrize the output, main text Eq.~(6)).

First, one can compute the transfer matrix for circuit in Fig.~\ref{fig:n_4_circ} as
\begin{equation}
  \frac{1}{2} \left( \begin{array}{cccc}
       1  & i & -1 & i \\
        i  & 1 & i & -1 \\
        -1  & i & 1 & i \\
        i  & -1 & i & 1 \\
    \end{array} \right),
\end{equation}
from which all relevant quantities can be calculated.

The output of the circuit output (pre-measurement), given four identical photons input on each rail $|1,1,1,1\rangle$ is
\begin{equation*}
\begin{split}
        & \frac{1}{2}|1,1,1,1\rangle + \\ 
        & \frac{1}{4}( |2,2,0,0\rangle -|2,0,2,0\rangle + |2,0,0,2\rangle  + \\
        & |0,2,2,0\rangle - |0,2,0,2\rangle + |0,0,2,2\rangle ) + \\ 
        &\sqrt{\frac{3}{32}}( |4,0,0,0\rangle + |0,4,0,0\rangle + |0,0,4,0\rangle + |0,0,0,4\rangle ).
\end{split}
\end{equation*}
We see the ideal output is now robust to detection errors to third order (e.g. requiring 2 dark counts as well as incorrect photon number resolution). In the absence of such errors, there is a 1/4 probability of detecting the desired output, and outputting a single photon in the unmeasured rail.

Moreover, one can calculate that in the event of a single error being sampled (each with probability $\epsilon (1-\epsilon)^3$), i.e., one of the $\{|\Psi_k\rangle \}_{k=1}^4$,
post-selection succeeds with probability $1/16$, of which $3/4$ of the outputted photons are the ideal $|\psi_0\rangle$. This follows from the observation that before measurement, in the subspace of a single photon per rail, the state is $\frac{1}{8}\sum_{k=1}^4 |\Psi_k\rangle$.

The measurement of a single photon per detected rail projects on to one of these four distinct outcomes, each with probability $1/64$. With total probability $3/64$ the outputted photon is ideal, and $1/64$ is an error state $|\psi_{i>0}\rangle$.
This is directly analogous to the 3 photon circuit.

Using this, one can bound the post-selection success probability by
\begin{equation*}
    P_{p.s.}^{(n=4)} \ge \frac{1}{4}(1-\epsilon)^3 = \frac{1}{4} - \frac{3}{4}\epsilon + O(\epsilon^2),
\end{equation*}
and the error reduction is
\begin{equation*}
    \epsilon' \le \frac{\epsilon}{4} + O(\epsilon^2).
\end{equation*}
By performing a more rigorous analysis as in App.~\ref{sect:eq_post_sel} (taking into account cases of more than a single error) we could achieve tighter bounds.

Overall this scheme requires, asymptotically, 16 photons to distill a single photon to $1/4$ the original error. 
Since it uses an additional beamsplitter, and requires an additional detection, it is likely the case that in a physical setting the $n=3$ circuit would be the best performing.

\section{Effect of Additional Errors \label{sect:errors}}
{The introduced distillation scheme has a built in robustness to detection errors since the Hamming distance between states in the ideal output (Eq.~(6) of the main text) is at least 2 over the measured rails; e.g. to mistake an outcome $|3,0,0\rangle$ or $|0,3,0\rangle$ with $|1,1,1\rangle$ by detection of the last two rails requires either two  dark counts, or a dark count and incorrect photon resolution respectively. 
If errors of this sort occur at a rate $\epsilon_d$, the overall contribution is $O(\epsilon_d^2)$, i.e. errors from faulty detections are only present as a second order effect. 
Moreover, even if one receives an error $|\psi_{i>0}\rangle$ in addition to a dark count, that could also result in a false measurement pattern, this is still a second order process $O(\epsilon \epsilon_d)$.}

In practice loss errors will also occur, which could mean no photon is present at the output, despite a successful detection pattern in the measured rails. In principle this is a first order effect, $O(\epsilon_L)$ where $\epsilon_L$ is the loss rate, since a single lost photon can yield the outcome $|0,1,1\rangle$ meaning the vacuum is outputted despite a valid detection pattern. It is important to note however that such errors will generally not build up, since as soon as more than one photon has leaked, we can not register a valid detection event (unless there are also dark counts, which again is a second order process). Moreover, these errors must occur at very specific locations in the circuit in order to be detrimental (for example loss in the measured rails is no problem, as the detection pattern will never be post-selected upon).

A similar discussion can be had on the converse side, where a photon source occasionally outputs two (or more) photons, instead of one. 
A simplistic model of this is where the photons are emitted at the same instance of time, still according to main text Eq.~(3).
If two photons are present at an input (i.e. $|2,1,1\rangle$ and permutations thereof), the post-selection only succeeds with probability 1/9, where two photons will be outputted by the circuit. Since the single-photon case nominally succeeds with probability 1/3,
these multi-photon errors will be filtered out at a rate of approximately $1/3$ per round. Or to say it another way, the two-photon sector of the post-selected state will decrease in magnitude by around $1/3$ per iteration. 

We also briefly mention that synchronization errors  -- where photons arrive at the inputs at slightly different times -- are expected to have relatively little effect on our scheme, so long as the intrinsic distinguishability of the source is still the dominant error in the system; erroneous photons will still be filtered out, though on average a greater number of photons would be required for distillation each round. The upper limit on the achievable fidelity will however depend on the effective distinguishability induced by synchronization errors.

\begin{figure}
    \centering
    \includegraphics[width=0.98\columnwidth]{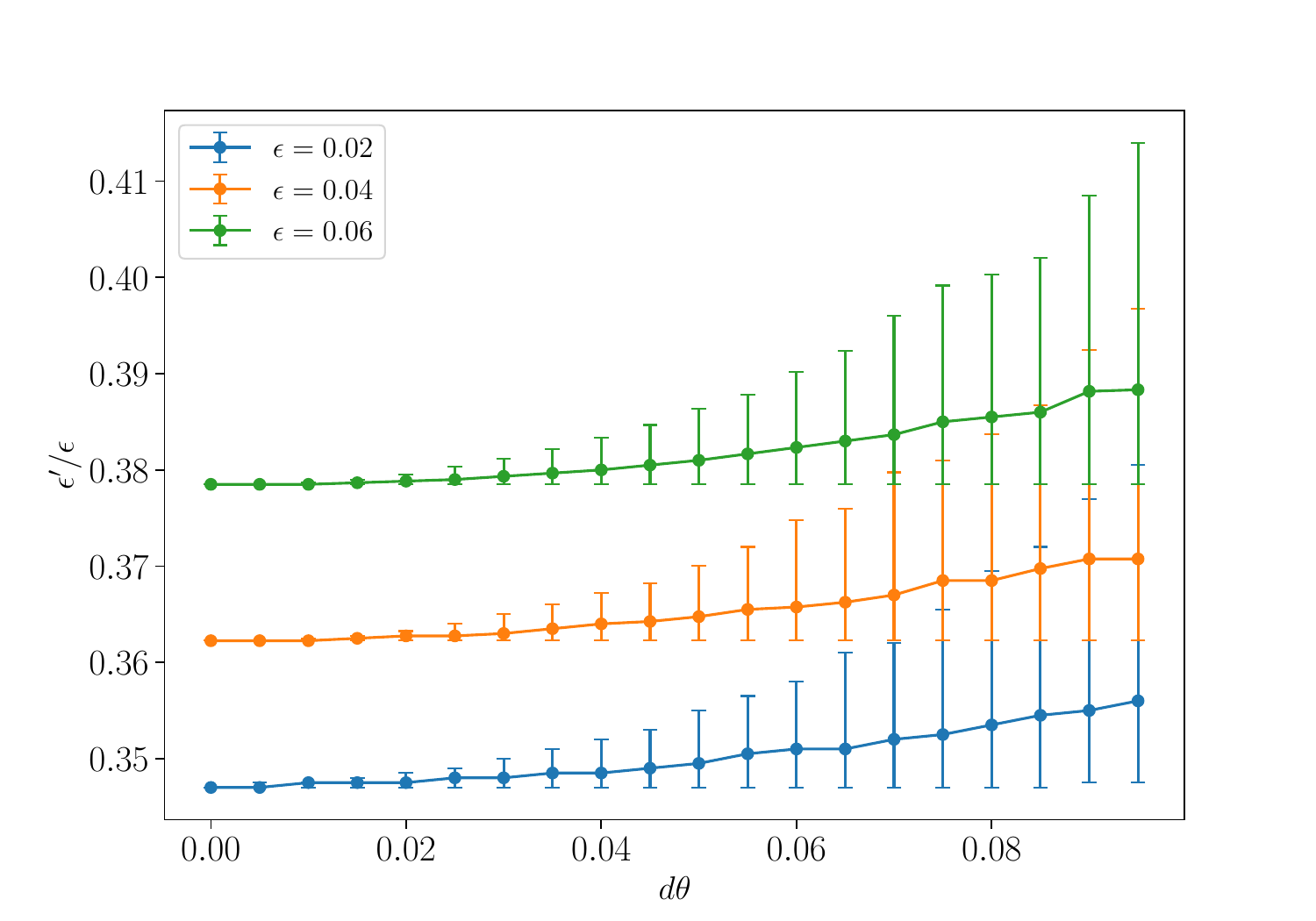}
    \caption{\textbf{Robustness in output fidelity under control errors.} For our $n=3$ scheme of the main text, we show the ratio of the output error $\epsilon^\prime$ to the input error $\epsilon$. The $d\theta$ quantity defines the magnitude of the control error on the beamsplitter angles, as described in the main text.
    For $d\theta=0$, we see this is close to $1/3$ as described in the main text for small $\epsilon$, but that the efficacy of the scheme decreases with increasing $d\theta$ (note, so long as $\epsilon' < \epsilon$, the scheme can be used to decrease the distinguishability). We plot the mean, and minimum and maximum of the quantity over 250 random samples drawn.}
    \label{fig:theta_epsilon}
\end{figure}

\begin{figure}
    \centering
    \includegraphics[width=0.98\columnwidth]{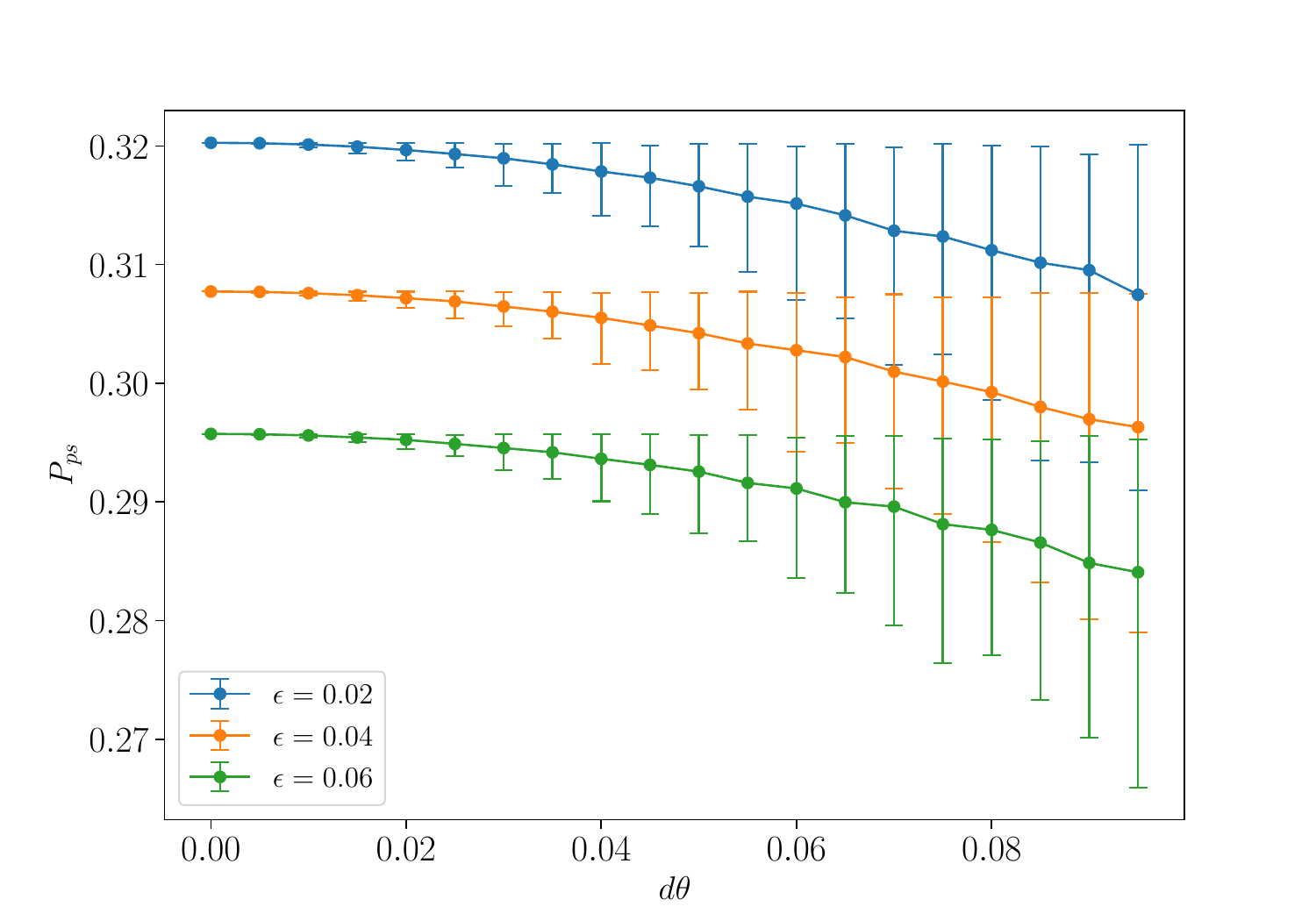}
    \caption{\textbf{Robustness in output post-selection probability under control errors.} For our $n=3$ scheme of the main text, we show post-selection success probability as a function of control error $d\theta$. The error $d\theta$ defines an error on the beamsplitter angles, as described in the main text.
    For $d\theta=0$, we see this is close to $1/3$ as described in the main text for small $\epsilon$, but that the post-selection probability decreases as the error in angles increases. This means more photons are required under the scheme (proportional to $1/P_{ps}$). We plot the mean, and minimum and maximum of the quantity over 250 random samples drawn.}
    \label{fig:theta_ps}
\end{figure}

{Finally, we show that the scheme is fairly robust to control errors from imperfect beamsplitter angles. In particular, if we wish to implement the angle $\theta$ which defines the transmission probability of a beamsplitter, but instead implement $\theta + d\theta$, we wish to understand the extent to which the present scheme can still be utilized. Interestingly, we find it has a high tolerance to such errors. In Figs.~\ref{fig:theta_epsilon}, \ref{fig:theta_ps}, we show the efficacy of the scheme under such errors, which indicates the scheme can still be used to improve the fidelity of a noisy photon source. In these simulations, for a given value of $d\theta$, we randomly update beamsplitter angles via $\theta \rightarrow \theta + R(-d\theta, d\theta)$, where $R(a, b)$ indicates a uniformly drawn random number in $(a, b)$. We do this for many samples to gather statistics (indicated by the error bars).}

{In these figures we see that whilst the efficacy of the scheme is generally reduced by the addition of a control error -- since the output fidelity $1-\epsilon^\prime$ and post-selection probability $P_{ps}$ are reduced -- both are still well within the regime where the scheme will be beneficial to use. Moreover, for control errors $d\theta \lesssim 0.05$, there is a minimal overhead in resources. For the angles required by our scheme ($\theta = \pi/4, \tan^{-1}\sqrt{2}$), this implies the scheme can easily tolerate up to around a 5\% error on the specified angles.}

\end{document}